\documentclass[twocolumn]{revtex4}%
\usepackage{amsfonts}%
\usepackage{amsmath}%
\usepackage{amssymb}%
\usepackage{graphicx}

\begin{document}
\title{Nonadiabatic geometric phase induced by a counterpart of the Stark shift }
\author{E. I. Duzzioni$^{1}$}
\email{duzzioni@df.ufscar.br}
\author{C. J. Villas-B\^{o}as$^{1}$, S. S. Mizrahi$^{1}$, M. H. Y. Moussa$^{1}$}
\author{R. M. Serra$^{1,2}$}
\email{serra@df.ufscar.br}
\affiliation{$^{1}$ Departamento de F\'{\i}sica, Universidade Federal de S\~{a}o Carlos,
P.O. Box 676, S\~{a}o Carlos, 13565-905, S\~{a}o Paulo, Brazil\linebreak$^{2}$
Optics Section, The Blackett Laboratory, Imperial College - London SW7 2BZ, UK}

\begin{abstract}
We analyse the geometric phase due to the Stark shift in a system composed of
a bosonic field, driven by time-dependent linear amplification, interacting
dispersively with a two-level (fermionic) system. We show that a geometric
phase factor in the joint state of the system, which depends on the fermionic
state (resulting form the Stark shift), is introduced by the amplification
process. A clear geometrical interpretation of this phenomenon is provided. We
also show how to measure this effect in an interferometric experiment and to
generate geometric \textquotedblleft Schr\"{o}dinger cat\textquotedblright%
-like states. Finally, considering the currently available technology, we
discuss a feasible scheme to control and measure such geometric phases in the
context of cavity quantum electrodynamics.

\textbf{Pacs Numbers:} 03.65.Vf, 42.50.Ct, 42.50.Pq

\textbf{Jornal Ref.} Europhys. Lett. \textbf{72}, 21 (2005)

\end{abstract}
\maketitle

Geometric phases have been studied more widely since the seminal\emph{\ }work
of Berry \cite{Berry}, in which he showed that a state, under an adiabatic and
cyclic evolution, acquires a phase of geometric origin that depends on its
path in the parameter space. This concept has been generalized in several ways
\cite{Shapere,Ben-Aryeh}, including noncyclic \cite{noncyclic}, nonadiabatic
\cite{nonadiabatic}, mixed state \cite{mixed} and open system \cite{open}
evolution. More recently the interest in geometric phases has grown, owing to
possible applications in quantum computation \cite{Vedral1}.

In the present paper, the geometric phase induced by a counterpart Stark shift
is investigated. We consider the dispersive interaction of a two-level
(fermionic) system with a quantized bosonic field driven by a time-dependent
(TD) linear amplification process. A nonadiabatic geometric phase factor in
the state of the system, which depends on the fermionic state, arises from the
TD linear amplification. This effect is due to distinct shifts in the field
frequency introduced by the different states of the two-level system (a
counterpart of the Stark shift) and it can be measured by an interferometric
experiment. We can interpret the origin of this phenomenon as a consequence of
the different projective maps associated with the dynamics of the two
fermionic states. Although, in general, the calculation of geometric phases in
the nonadiabatic case is not an easy task, the TD invariants technique of
Lewis and Riesenfeld \cite{Lewis} provides an easy and direct way to obtain
such phases \cite{Salomon,Mostafazadeh}.

We also propose a scheme, employing cavity quantum electrodynamics (QED) and
considering currently available technology \cite{Haroche1, Haroche2, Walther},
by which such phases can be generated, manipulated, and tested. Recent
advances in this context have led to striking experiments that afforded
fundamental tests of quantum theory \cite{Haroche1, Haroche2, Haroche3,
Walther}, as well as motivating several theoretical proposals. Carollo
\textit{et al. }\cite{Carollo1} proposed an experiment to measure the
adiabatic geometric phase induced by the vacuum state \cite{Carollo2} when a
two-level atom interacts resonantly with two quantized modes in a cavity. In
Ref. \cite{Serra1} the authors show, in a Jaynes-Cummings-like model, how to
simulate anyonic behavior and how to transmute the statistics of the
atom-field system via the adiabatic geometric phase.

Assuming that the transition frequency of the two-level system $\omega_{0}$ is
detuned enough from the bosonic field frequency $\nu$ \cite{Scully}, the
effective Hamiltonian for the dispersive interaction between the two-level
system and the field mode under TD linear amplification process, is given by
($\hbar=1$)
\begin{align}
\widehat{H}\left(  t\right)   &  =\nu\widehat{a}^{\dagger}\widehat{a}%
+\frac{\omega_{0}}{2}\widehat{\sigma}_{z}+\chi\widehat{a}^{\dagger}\widehat
{a}\widehat{\sigma}_{z}+\chi\widehat{\sigma}_{ee}\nonumber\\
&  +f(t)\widehat{a}^{\dagger}+f^{\ast}(t)\widehat{a}\text{,}\label{Eq1}%
\end{align}
where $\chi$ is the effective dispersive coupling constant
\cite{Scully,Haroche2}, $\widehat{\sigma}_{z}=\left\vert e\right\rangle
\left\langle e\right\vert -\left\vert g\right\rangle \left\langle g\right\vert
$ is the usual Pauli pseudo-spin operator, $\widehat{\sigma}_{ee}=\left\vert
e\right\rangle \left\langle e\right\vert $ ($\left\vert e\right\rangle $ and
$\left\vert g\right\rangle $ are the excited and ground states of the
two-level system, respectively\textbf{)}, $\widehat{a}^{\dagger}$
($\widehat{a}$) is the creation (annihilation) field operator and $f(t)$ is
the TD complex amplitude of the linear amplification.

The state vector of the Schr\"{o}dinger equation associated with Hamiltonian
(\ref{Eq1}) can be written as \cite{Celso}%
\begin{equation}
\left\vert \Psi(t)\right\rangle =e^{i\omega_{0}t/2}\left\vert g\right\rangle
\left\vert \Phi_{g}(t)\right\rangle +e^{-i\left(  \omega_{0}/2+\chi\right)
t}\left\vert e\right\rangle \left\vert \Phi_{e}(t)\right\rangle ,\label{Eq2}%
\end{equation}
with $|\Phi_{\ell}\left(  t\right)  \rangle=\widehat{1}_{a}\otimes\left\langle
\ell\right.  |\Psi\left(  t\right)  \rangle$ ($\ell=e,g$), $\widehat{1}_{a}$
being the unitary operator of field mode $a$ represented in a convenient
basis. Using the orthogonality of the fermionic states in $|\Psi\left(
t\right)  \rangle$ we obtain the uncoupled TD Schr\"{o}dinger equations for
the bosonic field states:%
\begin{equation}
i\frac{\partial}{\partial t}\left\vert \Phi_{\ell}(t)\right\rangle
=\widehat{\mathrm{H}}_{\ell}(t)\left\vert \Phi_{\ell}(t)\right\rangle
,\label{Eq3}%
\end{equation}
where%
\begin{equation}
\widehat{\mathrm{H}}_{\ell}(t)=\varpi_{\ell}\widehat{a}^{\dagger}\widehat
{a}+f(t)\widehat{a}^{\dagger}+f^{\ast}(t)\widehat{a},\label{Eq4}%
\end{equation}
with $\varpi_{e}=\nu+\chi,$ and $\varpi_{g}=\nu-\chi$. Now, the problem has
been reduced to that of a harmonic oscillator under linear amplification,
whose frequency $\nu$ is shifted by $\chi$ ($-\chi$) when interacting with the
state $\left\vert e\right\rangle $ ($\left\vert g\right\rangle $) \cite{Celso}.

We will employ the TD invariants technique to solve exactly Hamiltonian
(\ref{Eq4}) and obtain the geometric phase associated with the states
$\left\vert e\right\rangle $ and $\left\vert g\right\rangle $. From the
well-known theorem of Lewis and Riesenfeld (LR) \cite{Lewis} it follows that
the general solution of the TD Schr\"{o}dinger equation (\ref{Eq3}), given by
$\left\vert \Phi_{\ell}(t)\right\rangle =\sum_{m}c_{\ell,m}\exp\left[
i\phi_{\ell,m}(t)\right]  \left\vert m,t\right\rangle _{\ell},$ comprehends a
superposition of the eigenstates $\left\vert m,t\right\rangle _{\ell}$ of the
Hermitian invariant $I_{\ell}(t)$ ($dI_{\ell}(t)/dt\equiv\partial I_{\ell
}(t)/\partial t+i\left[  \widehat{\mathrm{H}}_{\ell}(t),I_{\ell}(t)\right]
=0$) multiplied by the TD phase factor $\phi_{\ell,m}(t)=\phi_{\ell,m}%
^{D}(t)+\phi_{\ell,m}^{G}(t)$; one of those components is dynamic:
\begin{equation}
\phi_{\ell,m}^{D}(t)=-\int_{0}^{t}dt^{\prime}\;_{\ell}\left\langle
m,t^{\prime}\right\vert \widehat{\mathrm{H}}_{\ell}(t^{\prime})\left\vert
m,t^{\prime}\right\rangle _{\ell},\label{Eq5d}%
\end{equation}
and the other is geometric \cite{Salomon,Mostafazadeh}:%
\begin{equation}
\phi_{\ell,m}^{G}(t)=i\int_{0}^{t}dt^{\prime}\;_{\ell}\left\langle
m,t^{\prime}\right\vert \frac{\partial}{\partial t^{\prime}}\left\vert
m,t^{\prime}\right\rangle _{\ell}.\label{Eq5g}%
\end{equation}
The geometric phase in this formulation arises from a holonomy over the
parameter space associated with the invariant $I_{\ell}(t)$
\cite{Salomon,Mostafazadeh}. By this method, it is possible to compute the
geometric phase in a general scenario, including adiabatic and nonadiabatic
evolutions of pure states.

The invariant $I_{\ell}(t)$ associated with Hamiltonian (\ref{Eq4}) is given
by \cite{Puri}%
\begin{equation}
I_{\ell}(t)=\widehat{a}^{\dagger}\widehat{a}-\alpha_{\ell}(t)\widehat
{a}^{\dagger}-\alpha_{\ell}^{\ast}(t)\widehat{a}+\beta_{\ell}(t),\label{Eq6}%
\end{equation}
and the invariant $\mathcal{I}(t)$ related to the total Hamiltonian
(\ref{Eq1}) is $\mathcal{I}(t)=%
{\displaystyle\sum\limits_{\ell=g,e}}
I_{\ell}(t)\left\vert \ell\right\rangle \left\langle \ell\right\vert ,$ where
the coefficients $\alpha_{\ell}(t)$ and $\beta_{\ell}(t)$ satisfy the coupled
equations
\begin{subequations}
\begin{align}
\overset{.}{\alpha_{\ell}}(t)  &  =-i\varpi_{\ell}\alpha_{\ell}%
(t)-if(t),\label{Eq7a}\\
\overset{.}{\beta_{\ell}}(t)  &  =i\alpha_{\ell}(t)f^{\ast}(t)-i\alpha_{\ell
}^{\ast}(t)f(t).\label{Eq7b}%
\end{align}

The eigenstates of the invariant $I_{\ell}(t)$ are the displaced number states
$\left\vert m,t\right\rangle _{\ell}=\widehat{D}\left[  \alpha_{\ell
}(t)\right]  \left\vert m\right\rangle $ where $\widehat{D}\left[
\alpha_{\ell}(t)\right]  =\exp\left[  \alpha_{\ell}(t)\widehat{a}^{\dagger
}-\alpha_{\ell}^{\ast}(t)\widehat{a}\right]  $ and $\alpha_{\ell}(t)$ is the
TD complex amplitude
\end{subequations}
\begin{equation}
\alpha_{\ell}(t)=\alpha_{\ell}(0)e^{-i\varpi_{\ell}t}-ie^{-i\varpi_{\ell}%
t}\int_{0}^{t}e^{i\varpi_{\ell}t^{\prime}}f(t^{\prime})dt^{\prime}.\label{Eq8}%
\end{equation}

Let us suppose that the fermionic particle is prepared initially in the
superposition state $c_{g}\left\vert g\right\rangle +c_{e}\left\vert
e\right\rangle $ while the field is in the vacuum state $\alpha_{\ell}(0)=0 $.
For our purpose we adjust the temporal dependence of the linear amplification
to resonance with the field mode, such that $f(t)=\kappa e^{-i\nu t}$. In
order to clarify the contributions of the geometric and the dynamic phases we
write the vector state of the system%
\begin{align}
\left\vert \Psi(t)\right\rangle  &  =e^{i\phi^{D}(t)}\left(  C_{g}%
(t)e^{i\phi_{g}^{G}(t)}\left\vert \alpha_{g}(t)\right\rangle \left\vert
g\right\rangle \right. \nonumber\\
&  \left.  +C_{e}(t)e^{i\phi_{e}^{G}(t)}\left\vert \alpha_{e}(t)\right\rangle
\left\vert e\right\rangle \right)  ,\label{Eq9}%
\end{align}
where $C_{g}(t)=c_{g}\operatorname*{e}\nolimits^{i\omega_{0}t/2}$,
$C_{e}(t)=c_{e}\operatorname*{e}\nolimits^{-i\left(  \omega_{0}/2+\chi\right)
t}$, and the amplitudes of the coherent states $\alpha_{\ell}(t)$ are given by%
\begin{equation}
\alpha_{\left(
\genfrac{}{}{0pt}{}{g}{e}%
\right)  }(t)=\pm\frac{\kappa}{\chi}e^{-i\nu t}(1-e^{\pm i\chi t}%
).\label{Eq10}%
\end{equation}
The geometric phases associated with the ground and excited states are
\cite{note1}
\begin{equation}
\phi_{\binom{g}{e}}^{G}(t)=\left(  2\nu\mp\chi\right)  \left(  \frac{\kappa
}{\chi}\right)  ^{2}\left[  t-\frac{\sin(\chi t)}{\chi}\right]  ,\label{Eq11}%
\end{equation}
while the factorized dynamic phase turns out to be%
\begin{align}
\phi^{D}(t) &  =\phi_{g}^{D}(t)=\phi_{e}^{D}(t)\nonumber\\
&  =-2\nu\left(  \frac{\kappa}{\chi}\right)  ^{2}\left[  t-\frac{\sin(\chi
t)}{\chi}\right]  .\label{Eq12}%
\end{align}
We note that, for the choice of the parameters considered above, the dynamic
phase remains factorized during the whole time evolution of the system, as
shown in Eq. (\ref{Eq9}), when the field mode is initially in the vacuum state
$\alpha_{\ell}(0)=0$; otherwise $\phi_{g}^{D}(t)\neq\phi_{e}^{D}(t)$. On the
other hand, the geometric phase depends on the fermionic state, i.e., on the
shift of the field frequency introduced by the fermionic particle.

Unlike the states of an isolated two-level system which can simply be
described on Bloch's sphere, in the present context a representation of the
geometrical phase associated with the evolution of the entangled state
(\ref{Eq9}) is supplied by getting rid of the fermionic degree of freedom and
making use of the quantum phase space associated with the field mode. From
this perspective, the dependence of the geometrical phase (\ref{Eq11}) upon
the state of the two-level system can be visualized in the projective map
$\Gamma_{\ell}$ of the Hilbert space $\mathcal{H}_{\ell}$ into the projective
space $\mathcal{P}_{\ell}$ [$\Gamma_{\ell}:\mathcal{H}_{\ell}\longrightarrow
\mathcal{P}_{\ell}$ ($\ell=e,g$)]. \textbf{\ }In fact, the Hamiltonian
(\ref{Eq4}) is different for each fermionic subspace, although the parameter
space is the same.

In Fig. 1 we present the typical spiraling path followed by the coherent
states $\left\vert \alpha_{\ell}(t)\right\rangle $, associated with the
fermionic levels $e$ and $g$, during a half cycle ($\chi t\in\lbrack0,\pi] $)
in the phase space $\operatorname{Re}\left\{  \left(  \chi/\kappa\right)
e^{i\upsilon t}\alpha_{\ell}(t)\right\}  \times\operatorname{Im}\left\{
\left(  \chi/\kappa\right)  e^{i\upsilon t}\alpha_{\ell}(t)\right\}  $. In
this figure, we can see that the trajectory of state $\left\vert \alpha
_{e}(t)\right\rangle $ performs an extra half loop in the phase space compared
to state $\left\vert \alpha_{g}(t)\right\rangle $, during a half cycle. This
occurs\textbf{\ }due to the different signals in Eq. (\ref{Eq10}) which arise
in the distinct frequency shifts of the effective Hamiltonian (\ref{Eq4}).
Summarizing, since the fermionic states introduce distinct shifts in the
bosonic field, the paths in the phase space associated to the respective
fermionic subspaces differ by one loop around the origin during one cycle.%

\begin{figure}
[h]
\begin{center}
\includegraphics[
trim=0.097592in 2.302805in 0.100032in 2.255918in,
height=6.1781cm,
width=8.0168cm
]%
{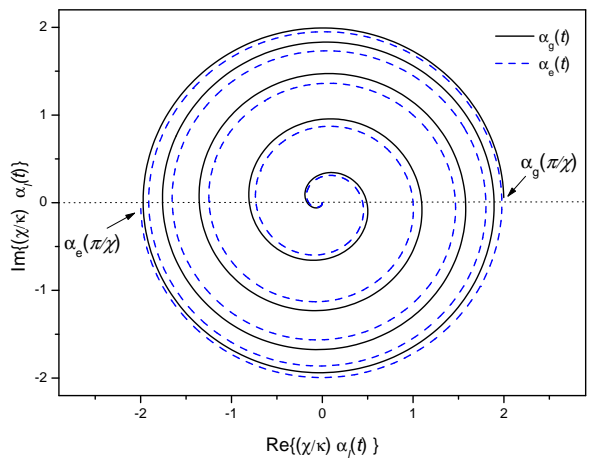}%
\caption{Typical path followed by the amplitudes $\alpha_{g}(t)$ and
$\alpha_{e}(t)$ of the coherent states in the phase space for $\chi
t\in\left[  0,\pi\right]  $.}%
\end{center}
\end{figure}

Let us consider the projective space associated with the instantaneous
eigenstates $\left\vert m,t\right\rangle _{\ell}$ of the invariant $I_{\ell
}(t)$. From the coupled equations (\ref{Eq7a}) and (\ref{Eq7b}) we can see
that the quantity $\beta_{\ell}(t)-\left\vert \alpha_{\ell}(t)\right\vert
^{2}$ (apart from a constant which can be considered null without loss of
generality) is conserved. Due to this conservation law we can represent the
state $\left\vert \alpha_{\ell}(t)\right\rangle $ in a surface of a
hyperboloid (i.e. the \textit{Poincar\'{e} hyperboloid }\cite{Ben-Aryeh}). In
Fig. 2 we show the trajectories of the state $\left\vert \alpha_{\ell
}(t)\right\rangle $ associated with each fermionic level $\ell$ in the
\textit{Poincar\'{e} hyperboloid}. In this figure we can clearly see that
there are two different projective maps $\Gamma_{\ell}$ associated with each
ferminonic state $\left\vert \ell\right\rangle $, introducing the effect shown
in Eq. (\ref{Eq11}).%

\begin{figure}
[th]
\begin{center}
\includegraphics[
trim=0.000000in 1.830735in 0.324495in 1.929838in,
height=7.3197cm,
width=8.2717cm
]%
{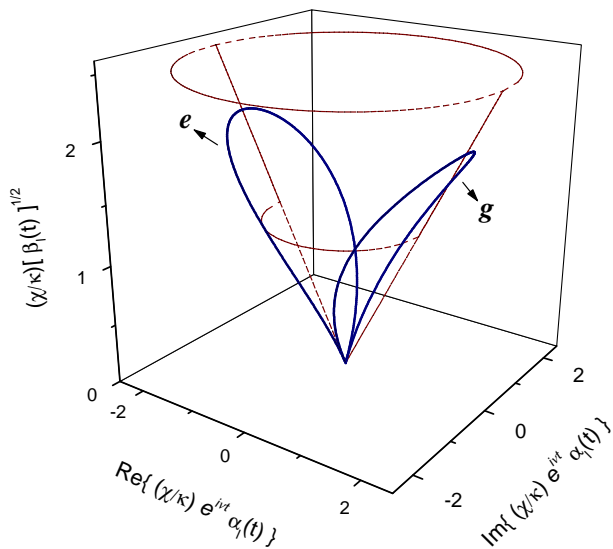}%
\caption{The path followed by the field states $\left\vert \alpha_{\ell
}(t)\right\rangle $ associated with each fermionic level ($g$ and $e$) in the
\textit{Poincar\'{e} hyperboloid }during a complete cycle (considering a
rotating frame with frequency $\nu$).}%
\end{center}
\end{figure}

Now, we will discuss how to implement, in a physical context, the
effect\textbf{\ }discussed above. To this end we consider the cavity QED
domain. The experimental setup proposed consists in a two-level Rydberg atom
which crosses a Ramsey-type arrangement (i.e., a high-Q microwave cavity $C$
placed between two Ramsey Zones $R_{1}$ and $R_{2}$) \cite{Haroche2} and is
detected in the excited state ($\left\vert e\right\rangle $) or in the ground
state ($\left\vert g\right\rangle $) by two ionization chambers $D_{e} $ or
$D_{g}$. The linear amplification in this context is achieved by the coupling
of a microwave generator to the cavity through a wave guide \cite{Haroche2}.

The geometric phase induced by a counterpart Stark shift can be verified in a
typical Ramsey-type interferometric experiment \cite{Haroche2}. Let us
consider\ the following scenario: first, the atom is prepared by a Ramsey zone
$R_{1}$ in a superposition state such that $c_{g}=c_{e}=1/\sqrt{2}$.
Subsequently, we assume that the microwave generator is turned on (off)
suddenly at the instant the atom enters (leaves) the cavity region\textbf{,}
so that $f(t)=0$ when the atom is outside the cavity. If the atom-field
interaction time is adjusted such that $\chi t=2\pi$, the evolved state vector
in the interaction picture assumes, apart from a global phase, the form
$\left\vert \Psi_{I}(t)\right\rangle =\left(  \left\vert g\right\rangle
+\operatorname*{e}\nolimits^{i\left[  \phi_{e}^{G}(t)-\phi_{g}^{G}(t)\right]
}\left\vert e\right\rangle \right)  \left\vert 0\right\rangle /\sqrt{2}$. In
the Ramsey zone $R_{2}$ a $\pi/2$ pulse [i.e. $\left\vert g\right\rangle
\longrightarrow\left(  \left\vert g\right\rangle +\left\vert e\right\rangle
\right)  /\sqrt{2}$ and $\left\vert e\right\rangle \longrightarrow\left(
\left\vert e\right\rangle -\left\vert g\right\rangle \right)  /\sqrt{2}$], is
performed in the atomic states. Finally the experiment is repeated for
different values of $\kappa$ to obtain the Ramsey fringes. In this way we can
measure the atomic state in the ionization chambers $D_{e}$ and $D_{g}$, so
that the atomic inversion (i.e. the difference between the detection
probability of the states $\left\vert e\right\rangle $ and $\left\vert
g\right\rangle $) becomes\textbf{\ }%
\begin{equation}
W_{eg}\left(  \kappa\right)  =\cos\left(  4\pi\frac{\kappa^{2}}{\chi^{2}%
}\right)  .\label{Eq13aa}%
\end{equation}
We note that in the present proposal the pattern of the Ramsey fringes depends
on the intensity $\kappa$ of the linear amplification field (for a fixed
coupling $\chi$) and not on the atom-field interaction time as usual
\cite{Haroche2}. The phase factor measured here is only of geometric nature.

In cavity QED experiments we have typically $\chi\sim10^{4}s^{-1}$ and
$\nu\sim10^{10}s^{-1}$ \cite{Haroche2}. Hence, it is possible to perform the
entire cycle in Fig. 2 assuming the interaction time $t\sim10^{-4}s$, which is
much shorter than the photon decay time --- of the order of $10^{-3}s$
\cite{Haroche1,Haroche2} for open and $10^{-1}s$ \cite{Walther} for closed
cavities --- making the dissipative and decoherence mechanisms practically
negligible. Therefore, our scheme provides a fast generation of geometric
phases, in contrast with the adiabatic proposals. We observe that to obtain an
argument of $2\pi$ in the \textit{cosine} in Eq. (\ref{Eq13aa}), we need
$\kappa=\chi/\sqrt{2}$ so that $\kappa\sim10^{4}s^{-1}$ for the above
parameters. In this situation the maximum amplitude of the coherent state
generated in the cavity mode, occurring in the half cycle, is $\left\vert
\alpha_{e}(t)\right\vert _{\max}=\left\vert \alpha_{g}(t)\right\vert _{\max
}=\left\vert 2\kappa/\chi\right\vert =\sqrt{2}$, so that the cavity mode has
less than two photons during the whole atom-field interaction. Another error
source is the imperfect synchronization between the switch on (off) of the
driving field and the instant when the atom enters (leaves) the cavity, due to
the spreading in the velocity of the atomic beam, which is typically (in
cavity QED experiments) about $2\%$ \cite{Haroche2}. Such spreading induces a
fluctuation in the time that the atom enters (leaves) the cavity about $\delta
t\sim10^{-6}s$, which leads to a small correction to the Eq. (\ref{Eq13aa}).
When the atom enters the cavity early we have%
\begin{equation}
W_{eg}^{early}\left(  \kappa\right)  \approx\cos\left[  4\pi\frac{\kappa^{2}%
}{\chi^{2}}-2\kappa^{2}\left(  \delta t\right)  ^{2}\right]  ,\label{Eq131}%
\end{equation}
and when the atom enters the cavity late the Eq. (\ref{Eq13aa}) turns out to
be%
\begin{equation}
W_{eg}^{late}\left(  \kappa\right)  \approx\cos\left[  4\pi\frac{\kappa^{2}%
}{\chi^{2}}-\kappa^{2}\chi\left(  \delta t\right)  ^{3}\right]  .\label{Eq132}%
\end{equation}
From the experimental parameters above, follows that the corrections in Eqs.
(\ref{Eq131}) and (\ref{Eq132}), due the imperfect synchronization, are
negligible $2\kappa^{2}\left(  \delta t\right)  ^{2}\sim10^{-4}$ and
$\kappa^{2}\chi\left(  \delta t\right)  ^{3}\sim10^{-6}$, respectively.

We stress that, the dispersive approximation, in the Hamiltonian
(\ref{Eq1}),\ can safely be assumed without significant corrections to the
computed geometric phase when considering the regime $\chi\geq\kappa\ $and
$\left(  \frac{g}{\Delta}\right)  ^{2}\ll1$ (where $\Delta=\nu-\omega_{0}$ is
the atom-field detuning and $g$ is the dipole Rabi coupling between the levels
$\left\vert e\right\rangle $ and $\left\vert g\right\rangle $). Such
conclusion follows from the analysis of the motion equations for the
transition operators $\left\vert e\right\rangle \left\langle g\right\vert $
and $\left\vert g\right\rangle \left\langle e\right\vert $ obtained without
the dispersive approximation. Typically, in the cavity QED experiments
\cite{Haroche2,Haroche3}, $\left(  \frac{g}{\Delta}\right)  ^{2}\sim10^{-3}$
and we have considered\ the parameter $\kappa$ satisfying the required regime.

The present interferometric device can also be employed to engineer
a\ \textquotedblleft Schr\"{o}dinger cat\textquotedblright-like state whose
parity depends on a geometric phase factor. If we adjust the interaction time
such that $\chi t=\pi$, the evolved state vector (after the atom has crossed
the Ramsey-type arrangement) turns out to be, apart from an irrelevant global
phase,%
\begin{align}
\left\vert \Psi_{I}(t)\right\rangle  &  =\frac{1}{2}\left[  \left(  \left\vert
2\kappa/\chi\right\rangle +\operatorname*{e}\nolimits^{2\pi i\left(
\kappa/\chi\right)  ^{2}}\left\vert -2\kappa/\chi\right\rangle \right)
\left\vert g\right\rangle \right. \nonumber\\
&  \left.  +\left(  \left\vert 2\kappa/\chi\right\rangle -\operatorname*{e}%
\nolimits^{2\pi i\left(  \kappa/\chi\right)  ^{2}}\left\vert -2\kappa
/\chi\right\rangle \right)  \left\vert e\right\rangle \right]  ,\label{Eq14}%
\end{align}
Therefore, the measurement of the atomic state projects the cavity mode into a
\textquotedblleft Schr\"{o}dinger-cat\textquotedblright-like state with a
relative phase of geometric nature.

It is worth mentioning, that the scheme presented here can be employed in
other experimental contexts.

In summary, we have shown that there is a geometric phase induced by the a
counterpart Stark shift and we have presented a scheme to control and measure
nonadiabatic geometric phases in cavity QED which can be carried out with
present-day technology. The dynamic phase in our scheme remains factorized
during the whole evolution of the joint system state and, on the other hand,
the geometric phases depend on the electronic states of the two-level atom.
This phenomenon arises from the Stark shift induced in the field mode by a
dispersive atom-field interaction. Finally, we note that this effect has a
potential application in geometric quantum computation \cite{Vedral1}, an
interesting topic for further investigation being the system composed by two
non-resonant atoms interacting dispersively with the cavity field under linear
amplification. In this case the shift in the field frequency will depend on
the joint state of the atoms and, in principle, by a suitable adjustment of
the parameters, a conditional operation could be obtained.

R. M. Serra acknowledge A. Carollo, D. K. L. Oi, F. L. Semi\~{a}o and V.
Vedral for enlightening discussions. This research was supported by FAPESP,
CNPq, and CAPES (Brazilian agencies).


\begin{thebibliography}{99}
\bibitem {note1}It is important to note that, the expression (13) agree with
the usual definition of the geometric phase for a noncyclic evolution, i.e.
$\ \arg\;_{\ell}\left\langle m,t^{\prime}=0\left\vert m,t^{\prime}=t\right.
\right\rangle _{\ell}-\operatorname{Im}\int_{0}^{t}dt^{\prime}\;_{\ell
}\left\langle m,t^{\prime}\right\vert \frac{d}{dt^{\prime}}\left\vert
m,t^{\prime}\right\rangle _{\ell}$ \cite{Ben-Aryeh,Mostafazadeh}, since the
first term is null due to our choice of initial state (the vacuum state).
\end{thebibliography}
\end{document}